\documentclass[preprint,showpacs,preprintnumbers,amsmath,amssymb,superscriptaddress]{revtex4}
\usepackage{amssymb}
\usepackage{amsmath}

\usepackage{graphicx}
\usepackage{txfonts}
\usepackage{subfigure}
\usepackage{dcolumn}
\usepackage{bm}

\begin{document}

\title{Time-dependent energetic laser-ion acceleration by strong charge separation field}

\author{Yongsheng Huang }
\email{huangyongs@gmail.com}
\author{Yuanjie Bi}
\affiliation{China Institute of Atomic Energy, Beijing 102413, China.}%

\affiliation{Department of Engineering Physics, Tsinghua University, Beijing 100084, China.}%
\author{Naiyan Wang}
\author{Xiuzhang Tang}
\affiliation{China Institute of Atomic Energy, Beijing 102413,
China.}
\author{Zhe Gao}
\affiliation{Department of Engineering Physics, Tsinghua University, Beijing 100084, China.}%



\date{\today}

\begin{abstract}
The laser-ion acceleration in the ultra-short and ultra-intense
laser-matter interactions attracts more and more interest nowadays.
Since electrons gain relativistic energy from laser pulse in a
period of several femtoseconds and driven away by the ponderomotive
force of laser pulse, a huge charge-separation field pulse is
generated. In general cases, the ion acceleration is determined by
this charge-separation field. A novel general time-dependent
solution for laser-plasma isothermal expansions into a vacuum with
different types of the scale length of the density gradient which
correspond to different charge separation forms is obtained. The
previous solutions are some special cases of our general solution. A
series of new solutions have been proposed and may be used to
predict new mechanisms of ion acceleration. However, many
unaccounted idiographic solutions that may be used to reveal new
acceleration mode of ions such as shock wave acceleration, may be
deduced from our general solutions.
\end{abstract}

\maketitle

\section{\label{sec:level1}Introduction}
The generation of energetic proton and acceleration mechanisms in
the ultra-intense laser pulses interaction with thin targets attract
more and more interest nowadays
\cite{Machnisms,M.Kaluza2004,Emmanuel
d'Humieres2005,Huang2007,HuangAPL1}. Their progress can provide
fundamental theory for inertial confined fusion (ICF) and promote
the realization of it effectively. The ultra-short and energetic ion
beam allows for an increase of energy resolution in the Time Of
Flight experiments, the investigation of the dynamics of nuclear
processes with high temporal resolution and the study of
spallation-related physics\cite{Spallation-related}.

When a relativistic laser pulse interacts with a plasma, the
laser-produced fast electrons with a unique temperature, $k_BT_e$,
determined by the laser ponderomotive potential are instantly
created and then driven away. However, the ions are still resting
due to the large mass and then a high charge separation field
generates. Furthermore, the plasma is assumed isothermal since the
continuous energy supply of the laser pulse in the pulse duration.
No matter proton shock acceleration (PSA)\cite{PSA1} in laser-plasma
interactions or target normal sheath acceleration (TNSA)\cite{S C
Wilks,P.Mora2,Huang2007,HuangAPL1} and so on, the ions are
accelerated by high charge-separation field. The key point of ion
acceleration is the hot-electron density distribution which decides
the spatial and temporal distribution of the charge-separation
field.

In this paper, a general solution for plasma isothermal expansions
into a vacuum is proposed with the assumption: the ion density
distribution can be represented a function with separable variables
in the transformation system. With the solution, the separate charge
distribution, electric field, electron velocity, ion velocity, and
fronts of ions and electrons are all predicted. For different
scaling length of the density, the solution corresponds to different
expansion mode of plasmas. In some special cases, the solutions have
been achieved by previous
pursuers\cite{P.Mora2,HuangAPL1,HuangAPL2}. A series of new special
solutions have been described and the corresponding acceleration
modes have been discussed in detail. It is pointed out the shock
wave forms for some types of the plasma density gradient and large
scale length.

\section{\label{sec:level1}Time-dependent ion acceleration due to strong charge separation}

\begin{figure}
{
\includegraphics[width=0.45\textwidth]{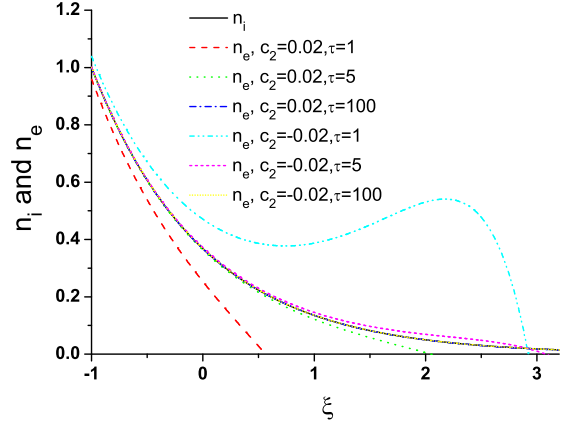}
}  \caption{\label{fig:ncom0} (Color online) The ion velocity,
$n_i$, and electron velocity, $n_e$ VS the self-similar variable,
$\xi$, for $L=-\beta_1=-1$ and $N_1\equiv1$ in two cases:
$u_0=c_2=0.02$ ; $u_0=c_2=-0.02$. }
\end{figure}
For convenience, the physical parameters: the time, $t$, the ion
position, $l$, the ion velocity, $v$, the electron field, $E$, the
electric potential, $\phi$, the plasma density, $n$, and the light
speed, $c$, are normalized as follows: $ \hat{t}={\omega_{pi0}t},
\hat{l}=l/\lambda_{D0}, u=v/c_s, \hat{E}={E}/{E_0},
\hat{\phi}=e\phi/k_BT_e, \hat{n}={n}/{n_{e0}}, \hat{c}=c/c_s, $
where $n$ represents $n_i$ (or $n_e$) which is the ion (or electron)
density, $n_{e0}$ is the reference hot-electron density,
$c_s=\sqrt{{Zk_BT_e}/{m_i}}$ is the ion acoustic speed,
$\omega_{pi0}=\sqrt{{Zn_{e0}e^2}/{m_i\epsilon_0}}$ is the initial
ion plasma frequency, $\lambda_{D0}=c_s\omega_{pi0}$, $c$ is the
light speed and $E_0={k_BT_e}/{e\lambda_{D0}}$. Here $e$ is the
elemental charge.


The reference frame used here is $\tau=\hat{t},
\xi=\hat{x}/\hat{t}$. With the transformation, the equations of
continuity and motion are obtained easily in the new coordinate
system. The ion density is assumed to satisfy: $
n_{i}(\tau,,\xi)=N_1(\tau)N_{2}(\xi)$. Here, $N_{1}\equiv1$
corresponds the self-similar ion density and solution. If
$N_2=exp(-\xi/\beta_1-1)$, where $\beta_1$ is a constant, the
self-similar solution is for a neutral-plasma isothermal expansion
into a vacuum given by Huang {\it et al.} \cite{HuangAPL1} for the
impurity ions. $\beta_1=1$ corresponds the classic self-similar
solution given by Mora in \cite{P.Mora2}. If
$N_2=n_0(\xi/\xi_0)^{-2/\alpha}$, where $\alpha\in (0,2)$, the
solution is for a half-self-similar non-neutral plasma isothermal
expansion into a vacuum proposed by Huang {\it et al.}
\cite{HuangAPL2}. For $dN_1/d\tau\neq 0$, the analytic solution has
not been reported. That is what given out by us next.

Combining the continuity and motion equation of ions gives the
general solution of the ion velocity and potential in the ion
region:
\begin{equation}\label{eq:uphi}
\begin{array}{c}
u_i=\alpha_2(\xi)-\delta_1(\tau)\alpha_1(\xi),\\
\phi=-\alpha_3^2/2+\int[\alpha_3+(\delta_1+\delta_2)\alpha_1]d\xi^{'},
\end{array}
\end{equation}
where $\delta_1=\tau/L_{\tau}$, $\delta_2=\tau^2/L_{\tau,2}$,
$L_{\tau}=[\partial \mathrm{ln}N_1/\partial \tau]^{-1}$ is the time
scale length of the ion density, $N_1(\tau)$,
$L_{\tau,2}=[\partial^2 \mathrm{ln}N_1/\partial \tau^2]^{-1}$,
$\alpha_1=F\int F^{-1}d\xi^{'}$ and $\alpha_2=F\int
F^{-1}\xi^{'}/Ld\xi^{'}$, $\alpha_3=\delta_1\alpha_1+\xi-\alpha_2$,
$F=\mathrm{exp}^{-\int d\xi^{'}/L}$ and $L=[\partial
\mathrm{ln}N_2/\partial \xi]^{-1}$ is the scale length of the
time-independent ion density, $N_2(\xi)$. And the electric field is
$E=-(\delta_1+\delta_2)\alpha_1/\tau-\alpha_3^2/L\tau+\delta_1\alpha_3/\tau$.

Combining Eq. (\ref{eq:uphi}) and Poisson's equation, the electron
density satisfies $n_e=n_i-\delta n$, where $\delta n$ is decided
by:
\begin{equation}\label{eq:ne}
\delta
n=\frac{2\alpha_3^2(1+L^{'}/2)}{L^2\tau^2}-\frac{(2+3\delta_1)\alpha_3-(\delta_1+\delta_2)\alpha_1-(\delta_1^2-\delta_2)L}{L\tau^2}.
\end{equation}
where $L^{'}=dL/d\xi$. From Eq. (\ref{eq:ne}),
$\lim_{\tau\rightarrow\infty}\delta n\rightarrow 0$ and it means
that the plasma tends to neutral as $\tau\rightarrow+\infty$,
whatever the initial state is.

We will confirm the ion front and electron front with physical
discussions and on the bases of Poisson's equation and the
continuity of the potential and electric field the next. The first
case is the electron front is beyond the ion front. Therefore,
beyond the ion front, the ion density is zero and it can be assumed
that the expression of electron density is a smooth expansion with
respect to self-similar variable, $\xi$. With the expression of
electron density, solving the continuity equation of electrons, the
electron density is $u_e=u_i+\delta u$, where $\delta u$ satisfies:
\begin{equation}\label{eq:ue}
\delta u=\frac{-\delta n
\alpha_3-(\partial\alpha_4/\partial\tau)/\tau+\alpha_4/\tau^2}{n_e}
\end{equation}
where
$\alpha_4=(\delta_1+\delta_2)\alpha_1+\alpha_3^2/L-\delta_1\alpha_3$
and then
$\tau\partial\alpha_4/\tau=(\delta_1-\delta_2-\delta_3+\delta_1^2+\delta_1\delta_2)\alpha_1+2\alpha_3(\xi-\alpha_2-\delta_2\alpha_1)/L+(\delta_2-\delta_1)\alpha_3$,
and $\delta_3=\tau^2\partial^3 N_1/\partial \tau^3$. In fact:
$\delta n=-(\partial\alpha_4/\partial\xi)/\tau^2$. With Eq.
(\ref{eq:ue}), $\lim_{\tau\rightarrow+\infty}u_e\rightarrow u_i$.

Therefore, with the electron density and Poisson's equation, the
electric field beyond the ion front satisfies:
\begin{equation}\label{eq:Ebey}
E(\tau,\xi)=-\tau
N_1\int^{\xi}_{\xi_{i,f}}N_2d\xi^{'}-\frac{\alpha_4(\xi)}{\tau},
\end{equation}
where $\xi_{i,f}$ represents the value of $\xi$ at the ion front. In
the ion region, the electric field is
$E(\tau,\xi)=-\alpha_4(\xi)/\tau$. From Eq. (\ref{eq:Ebey}),
$\xi_{i,f}$, satisfies:
\begin{equation}\label{eq:xiif}
N_1\int^{\xi_{e,f}}_{\xi_{i,f}}N_2d\xi^{'}=-\frac{\alpha_4(\xi_{e,f})}{\tau^2},
\end{equation}
where $\xi_{e,f}$ stands for the position of the electron front
there the electron density is zero.

With the expression of the ion front, $\xi_{i,f}$, maximum ion
velocity is given by:
\begin{equation}\label{eq:vimax}
u_{i,m}=\alpha_2(\xi_{i,f}|_{\tau_{acc}})-\delta_1\alpha_1(\xi_{i,f}|_{\tau_{acc}}),
\end{equation}
where $\tau_{acc}$ is the acceleration time, which is about $1-2$
times of the laser pulse duration for the ion acceleration in the
laser-solid interactions.

If the electron front is before the ion front, the ion velocity is
larger than that of electrons. However, in reality, this situation
can not happen. Therefore, we ignore the solutions in this case.

Before we discuss the different cases of the scale length, $L$, a
very special solution that does not rely on the choose of L is
given. If $N_1\propto1/\tau$, the special solution is:
\begin{equation}\label{eq:specialsol}
\begin{array}{c}
n_i=n_e=\frac{N_2(\xi)}{\tau}, u_i=u_e=\xi, \phi=\phi_0, E=0,
\end{array}
\end{equation}
This solution corresponds a neutral plasma with no charge separation
and a constant velocity expands into a vacuum.

In two special cases, the special solutions are familiar to us.

Case one: the scale length of ion density, L, is $0$-degree
polynomial in $\xi$, $L=-\beta_1$. If $L=-\beta_1$,
$F=\mathrm{exp}^{\xi/\beta_1+1}$, $N_2=F^{-1}$ and
$u_0=u(\xi=\xi_0)$ at $\xi_0=-\beta_1$. Then
$\alpha_1=-\beta_1(1-c_1F)$, $\alpha_2=\xi+\beta_1+c_2F$,
$\alpha_3=-\beta_1[1+\delta_1(1-c_1F)]-c_2F$ and
$u_i=\xi+\beta_1+c_2F+\beta_1\delta_1(1-c_1F)$, where $c_1$ and
$c_2$ are integral constants.

For $c_1=1$ and $c_2=0$, $u_0=0$. If
$\delta_1<\mathrm{exp}^{-\xi/\beta_1-1}$, ions are accelerated.
Oppositely, if $\delta_1>\mathrm{exp}^{-\xi/\beta_1-1}$, ions are
decelerated. Therefore, as Huang {\it et
al.}\cite{Huang2007,Huangunder} pointed out
$\delta_1\approx1>\mathrm{exp}^{-\xi/\beta_1-1}$ for
$\xi\geq\xi_0=-\beta_1$, this solution is not suitable to describe
the ion acceleration in the ultra-intense laser-foil interactions.

For $c_1=c_2=0$, a special solution: $\alpha_1=-\beta_1$,
$\alpha_2=\xi+\beta_1$, $\alpha_3=-\beta_1[1+\delta_1]$,
$u_i=\xi+\beta_1+\beta_1\delta_1$ and $\delta n=0$. Since
$\alpha_4=-\beta_1(1+\delta_2+2\delta_1)$, assuming
$\xi_{e,f}=+\infty$, from Eq. (\ref{eq:xiif}), the ion front is
\begin{equation}\label{eq:xiifL0c0}
\xi_{i,f}=\beta_1\mathrm{ln}(\frac{N_1\tau^2}{1+2\delta_1+\delta_2})-\beta_1.
\end{equation}
The main part of Eq. (\ref{eq:xiifL0c0}) is the similar as that
given by Huang and co-workers\cite{Huangunder} using physical
discussion. This solution is a special time-dependent solution for
neutral-plasma isothermal expansion into a vacuum given by Huang
{\it et al.}\cite{Huangunder}. In special case: $N_1\equiv1$, with
Eq. (\ref{eq:xiifL0c0}), the ion front is governed by:
$\xi_{i,f}=\beta_1[\mathrm{ln}({\tau^2})-1]$. Considering the
initial conditions, the results given here are the same as that
given by Huang {\it et al.}\cite{HuangAPL1} and Mora \cite{P.Mora2}
(where $\beta_1=1$). Huang {\it et al.} and Mora obtained the
results through the physical discussion about the Debye length of
electrons instead of analytic deductions. However, the same analytic
method has been used by Huang {\it et al.}\cite{HuangAPL2} to deduce
a special result of the following solution in case two.

For $N_1\equiv1$, $u_i=\xi+\beta_1+c_2F$. With Eq. (\ref{eq:ne}) and
Eq. (\ref{eq:ue}), $\delta n=-2c_2F(\beta_1+c_2F)/\tau^2$ and
$\delta
u=[(\beta_1+c_2F)^2(2c_2F+1/\beta_1)]/[\tau^2F^{-1}+2c_2F(\beta_1+c_2F)]$.
For $u_0=c_2=0$ ($\xi_0=-\beta_1$), $\delta n=0$ and the solution
corresponds the general self-similar solution for neutral-plasma
isothermal expansions into a vacuum pointed out by Huang {\it et
al.}\cite{HuangAPL1}. However, if $u_0=c_2>0$ ($\xi_0=-\beta_1$),
the acceleration described by the solution , which is not for a
neutral-plasma expansion, is more efficient than the classic
solution. If $u_0=c_2<-\beta_1$ ($\xi_0=-\beta_1$), the ion will be
accelerated in the opposite direction. If $-\beta_1<u_0=c_2<0$
($\xi_0=-\beta_1$), the ion will be accelerated first and
decelerated then. The density distributions for $c_2>0$ and $c_2<0$
have been shown by Fig. \ref{fig:ncom0}.

Fig. \ref{fig:ncom0} shows: $u_0=c_2=-0.02$ and the electron density
is smaller than the ion density for $\xi<3$ and the ion acceleration
is less efficient than that in the neutral-plasma case;
$u_0=c_2=0.02$ and the electron density is larger than the ion
density for all $\xi$ and the ion acceleration is more efficient
than that in the neutral-plasma case, since the hot-electron number
is so much more than that of ions. In these two cases, the density
difference becomes small with the increase of $\tau$ and tends to
zero as $\tau\rightarrow+\infty$. The electron front is determined
by: $u_e(\xi_{e,f})\approx \hat{c}$ in the relativistic laser
intensity. Then, with $\xi_{e,f}$, the ion front can be obtained by
Eq. (\ref{eq:xiif}).

For $c_2=0$, it is equivalent to:
$\xi_{e,f}+\beta_1+\beta_1\mathrm{exp}(\xi_{e,f}/\beta_1+1)/\tau^2]\approx
\hat{c}$. With this, the electron front satisfies:
$\xi_{e,f}<\hat{c}-\beta_1(1+\mathrm{exp}(1)/\tau^2)$. However, in
previous works given by Mora \cite{P.Mora2} and Huang and co-workers
\cite{HuangAPL1}, the electron front is taken as the positive
infinity. Then, with $\xi_{e,f}$ and Eq. (\ref{eq:xiif}), the ion
front can be obtained by:
$\xi_{i,f}=\beta_1\mathrm{ln}[\tau^2/(1+\beta_1^{'})]-\beta_1$,
where $\beta_1^{'}=\beta_1/(\hat{c}-\xi_{e,f}-\beta_1)$.

Case two: L is an one-degree polynomial in $\xi$, $L=-\alpha \xi/2$
and $\alpha\in(0,2)$. In this case,
$F=|\xi/\xi_0|^{2/\alpha}=\bar{\xi}^{2/\alpha}$,
$\alpha_1=-\beta\xi_0(\bar{\xi}+c_1F)$,
$\alpha_2=(1+\beta)\xi_0(\bar{\xi}+c_2F)$,
$\alpha_3/\xi_0=-\beta\bar{\xi}(1+\delta_1)-[\delta_1\beta
c_1+(1+\beta)c_2]F$, where $\beta=\alpha/(2-\alpha)$, $c_1$ and
$c_2$ are integral constants. With Eq. (\ref{eq:uphi}),
$u_i/\xi_0=\bar{\xi}[1+\beta(1+\delta_1)]+[\delta_1\beta
c_1+(1+\beta)c_2]F$. With this, the ion acceleration is determined
by the initial state and the density distribution (neutral or not).

\begin{figure}
{
\includegraphics[width=0.45\textwidth]{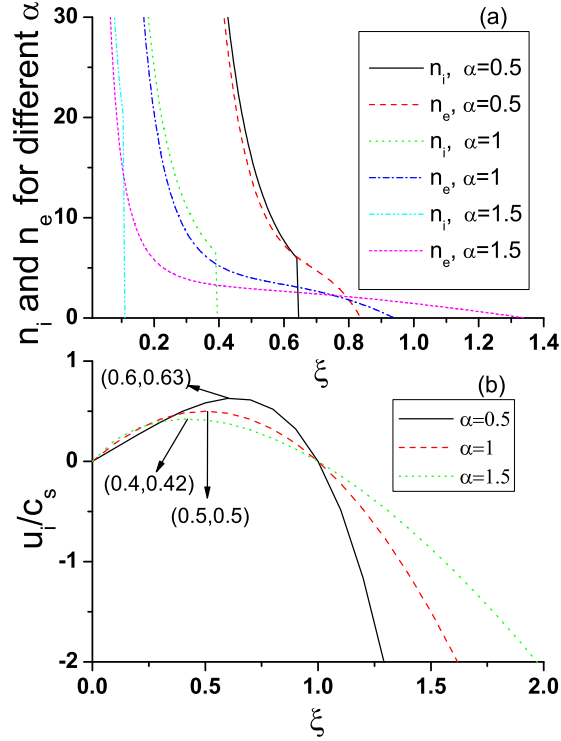}
}  \caption{\label{fig:vncomL1order} (Color online) (a) The electron
and ion density VS the self-similar variable, $\xi$, for
$L=-(\alpha/2)\xi$ in three cases: $\alpha=0.5,1,1.5$. The electron
fronts are all before the ion fronts. (b) The ion velocity VS the
self-similar variable, $\xi$, for $L=-(\alpha/2)\xi$ and
$\alpha=0.5,1,1.5$. In Fig. \ref{fig:vncomL1order}(a) and (b),
$u_0(\xi=\xi_0)=0$, $c_1=c_2=-1$, $\xi_0=1$, $N_1\equiv1$. }
\end{figure}

For $u_0(\xi=\xi_0)=0$, $c_1=c_2=-1$, and the ion velocity is
$\bar{\xi}\xi_0[1+\beta(1+\delta_1)](1-\bar{\xi}^{1/\beta})<0$ for
$\xi>\xi_0>0$, $u_i=0$ for $\xi=\xi_0$ , and $u_i>0$ for
$0<\xi<\xi_0$. All the ions move to the central point: $\xi=\xi_0$,
and the farther the distant, the larger the ion speed. With Eq.
(\ref{eq:ne}), the density difference, $\delta n$, is calculated for
$N_1\equiv1$. In this case, the electron front is given by $\delta
n(\xi_{e,f})=\bar{\xi_{e,f}}^{-2/\alpha}$. With Eq. (\ref{eq:xiif}),
the ion front is obtained.  Fig. \ref{fig:vncomL1order}(a) shows the
ion and electron density at $\tau=1$ and $\alpha=0.5,1,1.5$. The ion
and electron front are all obtained. This situation is not an
efficient acceleration mode since maximum ion velocity for any
$\alpha\in(0,2)$ is very finite as shown by Fig.
\ref{fig:vncomL1order}(b).

\begin{figure}
{
\includegraphics[width=0.5\textwidth]{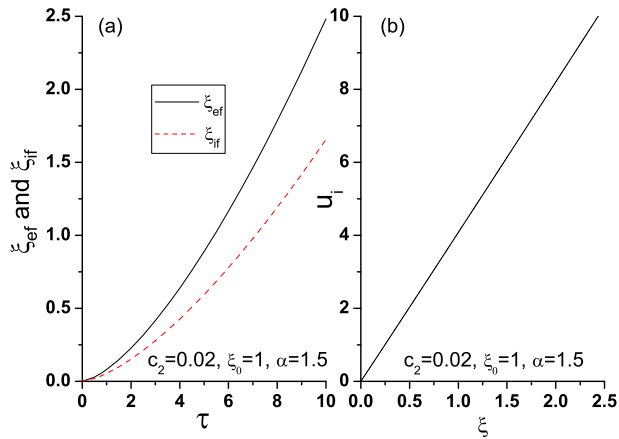}
}  \caption{\label{fig:frontuicomL1} (Color online) (a) The ion and
electron front VS the expanding time, $\tau$; (b) the ion velocity
VS the self-similar variable, $\xi$, for $L=-\beta\xi/2$ and
$N_1\equiv1$. Here, $c_2=0.02$, $\alpha=1.5$ and $\xi_0=1$. The ion
is accelerated efficiently and about $6$ times acoustic velocity at
$\tau=10$. }
\end{figure}

However, if $c_2=0.02$ and $N_1\equiv1$, the ion velocity is
$\bar{\xi}\xi_0[1+\beta(1+\delta_1)](1+0.02\bar{\xi}^{1/\beta})$.
Fig. \ref{fig:frontuicomL1} shows the ion acceleration is efficient
in this case, especially for $\alpha=1.5$ and $\xi_0=1$. The
electron and ion front are obtained with $n_e(\xi_{e,f})=0$ and Eq.
(\ref{eq:xiif}) and shown by Fig. (\ref{fig:frontuicomL1})(a).

For $c_1=c_2=0$, $u_i=\xi+\beta\xi(1+\delta_1)$and
$\phi=-(1+\beta)\beta\xi^2(1+\delta_1)^2/2-(\delta_2-\delta_1^2)\beta\xi^2/2$.
$\delta
n=[\beta\delta_2+\beta^2\delta_1^2+\beta(1+\beta)(2\delta_1+1)]/\tau^2$
This solution is also time-dependent and can be used to describe the
energetic ion acceleration with an enhanced electron tail. For
$N_1\equiv1$, the solution has been obtained by Huang {\it et al.}
\cite{HuangAPL2}.
$\alpha_4=-[\delta_2+\beta(1+\delta_1)^2+1+2\delta_1]\beta\xi$ and
$\xi_{i,f}=(\alpha/2)^{\beta}\xi_{e,f}$, and the electron front is
given by:
\begin{equation}\label{eq:xefnon}
\xi_{e,f}=\{\frac{N_1n_0}{[\delta_2+\beta(1+\delta_1)^2+1+2\delta_1]\beta}\}^{\alpha/2}\tau^{\alpha}.
\end{equation}
In special case: $N_1\equiv1$, the results from Eq.
(\ref{eq:xefnon}) are the same as that obtained in \cite{HuangAPL2}.

Two special cases: L are zero-degree and one-degree polynomials in
$\xi$ have been calculated and some time-independent solutions are
the same as the previous works. For different L, the ion
accelerations are different. The key variable is the scale length of
the time-independent electron density, $L$. In the neutral-plasma
case, $L=-\beta_1$ ($\beta_1\in(0,1)$), which is a zero-degree
polynomial in $\xi$. In the hot-electron-tail case,
$L=-(\alpha/2)\xi$ ($\alpha/2\in(0,1)$), which is a one-degree
polynomial in $\xi$. Even for the same L, different integral
constants induce different solutions. However, the essential
determinant is the charge separation: $\delta n$. Therefore, it is
concluded that the charge separation of a plasma determines the ion
acceleration.

Here, the case: $L$ is a quadratic polynomial in $\xi$,
$-\beta_2\xi^2$ ($\beta_2\in(0,1]$), is considered the first time.
For a n-degree polynomial: $L=-\beta_n\xi^n$ can be considered in
the same way.

If $L=-\beta_2\xi^2 (\beta_2\in(0,1])$ and $N_1\equiv1$,
$F=\mathrm{exp}(-1/\beta_1\xi+1/\beta_1\xi_0)$, $n_i=F^{-1}$ and
$u_i=\alpha_2=F\int_{\xi_0}^{\xi}F^{-1}d\xi^{'}+u_0F$, where
$u_0=u_i(\xi=\xi_0)$ is a constant. With this,
$\alpha_3=\xi-\alpha_2$, $\alpha_4=-(1-\alpha_2/\xi)^2/\beta_2$,
$\delta
n=2(1-\alpha_2/\xi)(1+\beta_2\alpha_2-\alpha_2/\xi)/\beta_2^2\tau^2$.
Since $\lim_{\xi\rightarrow+\infty}\alpha_2\rightarrow
\xi+u_0\exp(1/\beta_2\xi_0)$, some results are obtained:
$\lim_{\xi\rightarrow+\infty}u_i\rightarrow
\xi+u_0\exp(1/\beta_2\xi_0)$,
$\lim_{\xi\rightarrow+\infty}\alpha_3\rightarrow0$,
$\lim_{\xi\rightarrow+\infty}\alpha_4\rightarrow0$ and
$\lim_{\xi\rightarrow+\infty}\delta n\rightarrow0$,
$\lim_{\xi\rightarrow+\infty}n_i\rightarrow n_e$. With Eq.
(\ref{eq:xiif}), the ion front and the electron front are all
positive infinity. However, $\partial u_i/\partial \xi<0$ for any
$u_0$ and $\xi$ large enough. Therefore, the shock wave forms for
$\xi$ large enough in this case.

Similar with above discussion, for $L=-\beta_n\xi^n$, the physics of
the expansions  of plasmas can be obtained easily and the same
results of them are: the plasma front is positive infinity and the
boundary condition should be added exteriorly. In fact, it is easy
to prove that the analytic form of the solutions can not be obtained
and they may correspond to shock section for $L=-\beta_n\xi^n,
n\geq2$ and $\xi$ large enough.

It is required that $\partial L/\partial\xi<0$ and $L\neq
-\beta_n\xi^n, n\geq2$ in order to make the limitation of the plasma
density be finite as $\xi$ tends to positive infinity. $L=-\beta_1$
and $L=-\alpha\xi/2$ are two simple cases.

In order to show time-dependent solutions, it ia assumed that
$N_1=\kappa \tau\propto \tau, \tau\in[0,\tau_u]$ as an example in
two cases: $L=-\beta_1$ and $L=-\alpha\xi/2$. $N_1=\kappa\tau$, then
$\delta_1=1$ and $\delta_2=0$. For $L=-\beta_1$ and $c_1=c_2=0$,
$u_i=\xi+2\beta_1$ and
$u_{i,f}=\beta_1\mathrm{ln}({\kappa\tau^3}e/{3})$, where
$e=2.718...$. For $L=-\alpha\xi/2$ and $c_1=c_2=0$,
$u_i=(1+2\beta)\xi$ and $\xi_{e,f}={[\kappa
n_0/(4\beta+3)\beta]}^{\alpha/2}\tau^{3\alpha/2}$. Considering the
dependence of the electron density on $\tau$ described by Huang {\it
et al.}in \cite{Huang2007}, these solutions can be used to describe
the influence of the hot-electron recirculation on the ion
acceleration at the rear of the target heated by ultra-intense laser
pulse. Similar with above discussions, the ion front and electron
front in the general case can also be obtained. Here, it is not
repeated again.

\section{\label{sec:level1}Conclusion}
In conclusion, a general time-dependent isothermal expansion for the
ion acceleration due to charge separation is proposed on the base of
the equations of continuity and motion of ions and Poisson's
Equation. As examples, several new solutions for each types of the
plasma density gradient have been proposed. Especially, for
$L=-\beta_n\xi^n, n\geq2$ and $\xi$ large enough, we pointed out
that the shock wave solution exists. However, the analytic formation
can not be achieved easily here.

This work was supported by the Key Project of Chinese National
Programs for Fundamental Research (973 Program) under contract No.
$2006CB806004$ and
 the Chinese National Natural Science Foundation under contract No.
$10334110$.



\end{document}